# Tayloring neutron optical performance of Fe/Si multilayers

**D.G. Merkel[1], B. Nagy[1], Sz. Sajti[1], E. Szilágyi[1], R. Kovács-Mezei[2] and L. Bottyán[1]**

[1]KFKI Research Institute for Particle and Nuclear Physics, P.O.B 49, H 1525, Budapest, Hungary
[2]Mirrotron Ltd. Konkoly-Thege M. út 29-33; H-1121 Budapest, Hungary

E-mail: merkel@rmki.kfki.hu

**Abstract.** Reflected neutron beam originating from a crystal monochromator contains higher order wavelength contribution. Creating multilayer mirror structures with custom reflectivity curves including a monochromatic polarized neutron beam is a challenge in modern neutron optics. In this work we present the study of three types of magnetron-sputtered aperiodic Fe/Si layer structures with the purpose of higher harmonic suppression. First, an approximate sinusoidal profile was achieved directly by carefully controlling the evaporation parameters during sputtering process. Second, we implemented a random distribution of the layer thicknesses in which the layer structure of the sample was derived from a prescribed simulated spectrum. Third, a quasi-sinusoidal rounded bilayer scattering length profile was attempted to achieve by 0, 5, 10, 27 and $270 \times 10^{15}$ /cm$^2$ fluence of 350 keV neon irradiation starting from a sharp bilayer profile. Besides the highest fluence, which considerably decreased the overall bilayer contrast, we found an improvement of the monochromatisation with increasing irradiation fluence by the decrease of the total intensity of higher order reflections from 11.1% to 2.2% while the intensity of the first order Bragg peak decreased from 80% to 70%. The polarizing efficiency also increased with increasing fluence of Ne+ irradiation from 78.8% to 90.7%. We were able to achieve 92% and 82% reflectivity of the random multilayer structure and sinusoidal profile sample with 87.6% and 97.1% polarizing efficiency, respectively.

## 1. Introduction

Multilayer mirrors are indispensable devices in slow neutron research. Periodic multilayers consisting of two materials with different neutron optical potential can be used as a neutron monochromator [1,2]. Due to their high scattering length density contrast, nickel and titanium are widely used in such devices [3,4]. However, to produce a polarized monochromatic beam, a series of alternating ferromagnetic (Fe, Co) and nonmagnetic layers (Ge, Si, etc) are used instead [578]. The main problem with using multilayer systems for monochromator is the presence of higher order reflections, which are to be rejected in most applications. As an alternative to higher harmonics filtering, Padiyath et al. [9] published a suitable sinusoidal-like modification of the mirror's bilayer depth profile in the case of the Ni/Ti system. The reflectivity curve being the refraction-corrected Fourier transform of the layers scattering length density profile, a single Bragg peak monochromator structure is of sinusoidal bilayer profile. In general, the preparation of a perfect sinusoidal layer profile in industrial scale (the case for neutron mirrors) is rather lengthy and cumbersome. Alternative methods of suppression of the intensity of higher order reflections can be beneficial. In this work we present three different approaches to decrease the higher harmonic reflection in magnetic (Fe/Si) multilayer monochromators. In the first method an approximately sinusoidal profile was attempted to achieve directly by carefully controlling the evaporation parameters during sputtering process. In the second instant, we implemented a random distribution of the layer thicknesses where the sample layer structure was derived from a prescribed simulated spectrum. Thirdly, the rounded quasi-sinusoidal scattering length density (SLD) profile was achieved by applying various fluences of 350 keV Ne+ irradiation. Rutherford backscattering spectrometry (RBS) and polarized neutron reflectometry (PNR) measurements were carried out to follow the effect of irradiation and PNR to study optical properties of the Fe/Si multilayer samples.

## 2. Experimental

Three different types of multilayers consisting of alternating Fe and Si layers were deposited on float glass using DC magnetron sputtering [10].During the growth process of silicon the pressure of the working gas (Ar) and the sputtering and power was kept at $2.6\times10^{-3}$ mbar and 500 W as compared to the case of iron with $1.3\times10^{-3}$ mbar and 750 W, respectively.

Since the evaporation of layers with very small layer thickness and a perfect control of coevaporaton of hundreds of layers is rather difficult, alternative layer sequences for higher order intensity suppression were constructed which can be produced on industrial scale sputtering machines. The concept of samples with sinusoidal layer profile is demonstrated in Figure 1.

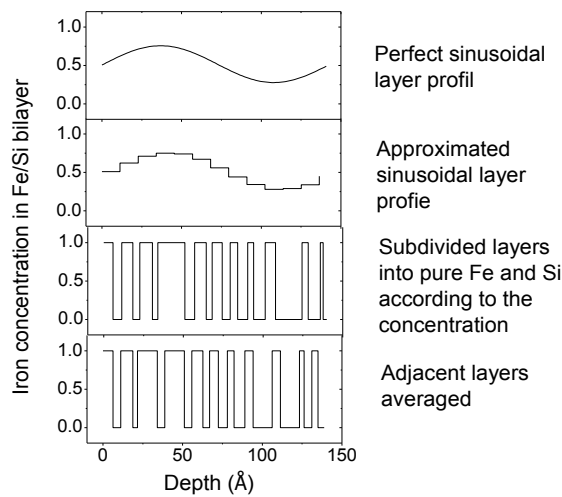

**Figure 1** The steps of creation of a sinusoidal-like layer profile

As shown in Figure 1 starting from a perfect sinusoidal iron concentration within an Fe/Si bilayer, the profile was approximated by a step-function. At a given concentration the mixed layer was subdivided into two pure iron and silicon layer with bilayer thicknesses corresponding to the concentration. In order to decrease the layer thickness variations the thickness of the adjacent layers were then averaged. Two samples; one with 138.5 Å bilayer thickness containing 22 sublayer repeated by 70 (Sample "sin1") and one with 143.1 Å bilayer thickness with 18 sublayers repeated 40 times (Sample "sin2") were prepared using the latter method.

In another approach the layer structure was derived starting from a random layer thickness sequence by fitting the layer structure to the desired neutron reflectivity curve as if they were measured data using Fitsuite [11]. The resulting Fe/Si layer structure consisted of 69 layers, starting and finishing with silicon (Sample "rand").

Finally, a nominally perfect $[Fe/Si]_{12}$ multilayer system with thicknesses 84 and 109 Å, respectively was prepared (Sample "MLS"). This latter sample was cut into pieces and the different pieces were irradiated by 350 keV $Ne^+$ ions of fluences $(1-27)\times10^{15}$ /cm$^2$. The energy was calculated (using the SRIM code [12]) so that the impinging ions would penetrate deep into the float glass substrate, hence the energy of the neon ions can be considered constant in the multilayer. The sample composition of the implanted samples and lateral inhomogenity of the "full" sample (with a size of $100\times250$ mm$^2$) was also determined by RBS using 1620 keV $He^+$ beam at a scattering angle of 165° and tilt angles of 7°, 70°and 80°. The full sample was found to be homogenous within ±10%. Significant mixing has been found only at the highest implanted fluence $27\times10^{15}$/cm$^2$.

The initial layer profiles for all four kind of samples are shown in Figure 2.

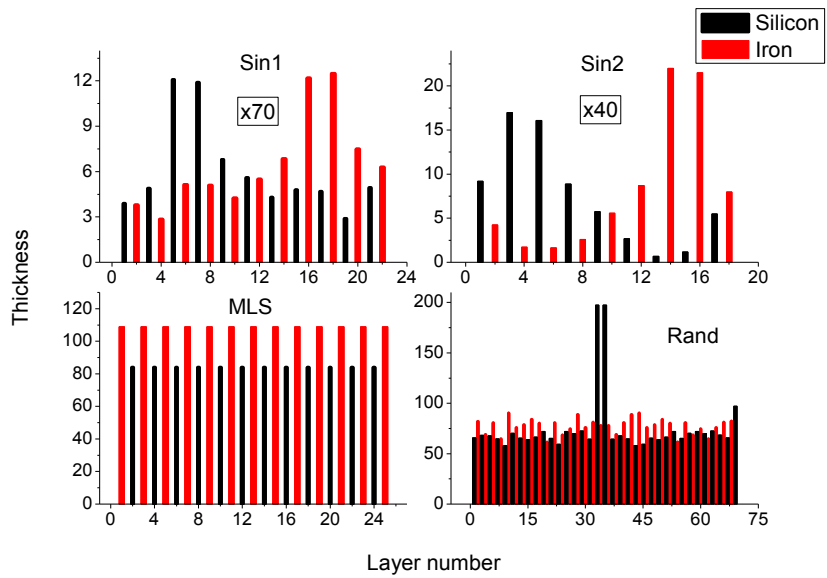

**Figure 2** Layer structure of the different samples in this study

In order to follow the changes in the neutron optical properties upon irradiation, polarized neutron reflectivity curves at wavelength of 4.59 Å with up and down polarization were recorded on the GINA reflectometer at the Budapest Neutron Centre, equipped with a supermirror polarizer and radio frequency adiabatic spin flipper. [13]

## 3. Results and discussion

PNR measurements of samples with approximate sinusoidal layer structure:

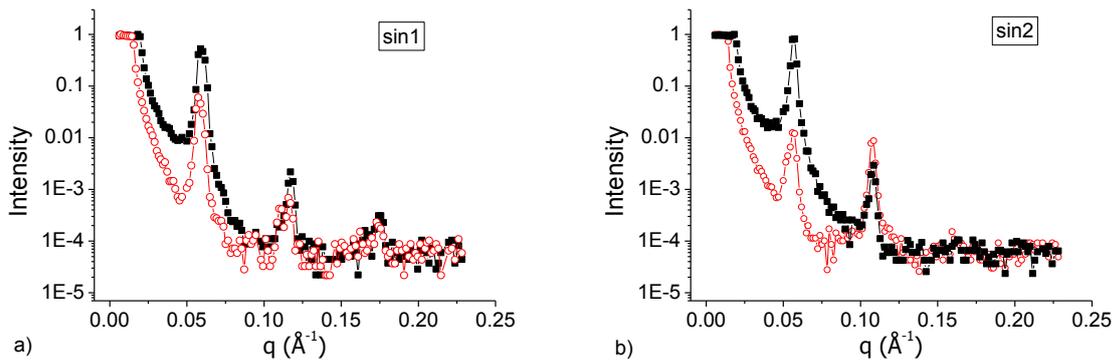

**Figure 3** Polarized neutron reflectivity ($R^+$ (Black) and $R^-$ (Red)) of two samples of approximate sinusoidal layer structure

Figure 3 a) shows the spectrum of sample "sin1" where the bilayer of 138.5 Å thickness was divided into 22 sublayers. The intensity of the first Bragg peak (in up polarization) was found to be 53% compared to the total reflection intensity which is characteristic of imperfection in the periodicity. At higher $q$ values, the higher order harmonics (n=2, 3) are also presented with 0.2% and 0.03%. The polarizing efficiency was calculated according to the following equation [14]:

$$P = \frac{I^+ - I^-}{I^+ + I^-} \qquad 1)$$

where $I^+$ and $I^-$ denote the intensity at a given $q$ value in up and down polarizations, respectively. Using the latter formula the polarization efficiency at the first order Bragg peak was found to be 79%.

In Figure 3 b) the PNR curve taken from sample "sin2" (bilayer thickness 148.1 Å divided into 18 sublayer) is shown. As opposed to sample "sin1" the first order Bragg peak - in up polarization - is significantly larger (82%) and among the harmonics, only n=2 order reflection appear with 0.29%. With this type of layer structure, a polarization efficiency of 97.1% was be achieved.

The PNR of the third sample (rand) is shown in Figure 4. The inset shows the simulated reflectivity of the original structure from which the layer structure of this sample was derived.

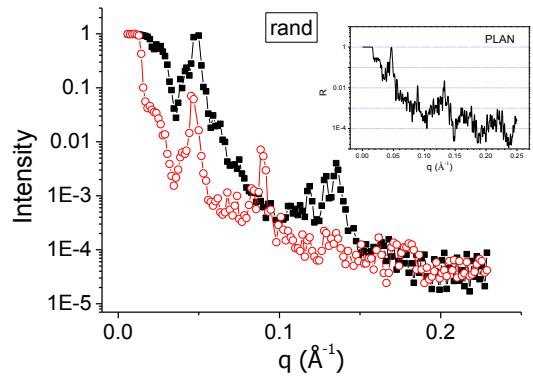

**Figure 4** PNR measurement ($R^+$ (Black) and $R^-$ (Red)) taken on sample which structure was determined by finding the proper layer profile the desired PNR simulation. Inset: the simulated spectrum

By examining the up curve, it can be seen that the n=2 and n=4 Bragg peaks are missing and it contains only two Bragg peaks (n=1 and n=3) with normalized intensities of 92% and 0.4%. In the other polarization, however, the first three Bragg peaks are present. The calculation of the polarizing efficiency yields 87%.

In order to follow the effect of 350 keV NE+ irradiation to the neutron optical properties of a Fe/Si multilayer structure (MLS) PNR measurements were carried out. The footprint-corrected PNR curves are plotted in Figure 5.

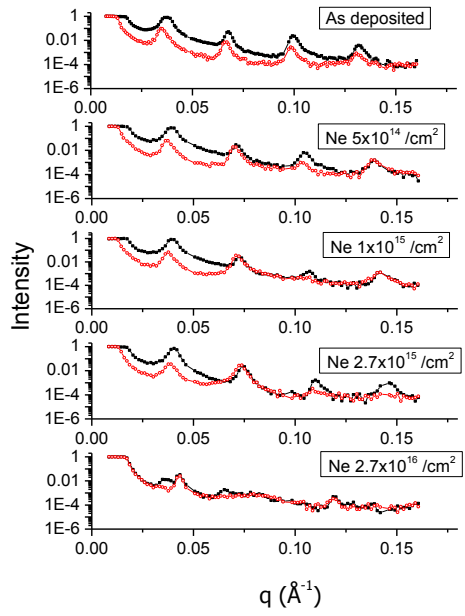

**Figure 5** PNR curves ($R^+$ (Black) and $R^-$ (Red)) recorded on Fe/Si multilayer samples after 0, 5, 10, 27 and 270×$10^{15}$ /cm² 350 keV Ne+ irradiation

At the as deposited sample Bragg peaks up to 4$^{th}$ order are detected in both up and down polarizations. The analysis of the Bragg peak intensities show constant decrease of higher harmonics followed by the decrease of the first order Bragg peak intensity from 80% to 70% up to a fluence of $2.7\times10^{16}$ /cm² where the intensity dropped to 3%. At this fluence the spectrum indicates a very high level of mixing in the adjacent layers. The detailed value of Bragg peak intensities together with the previous samples are plotted in Figure 6.

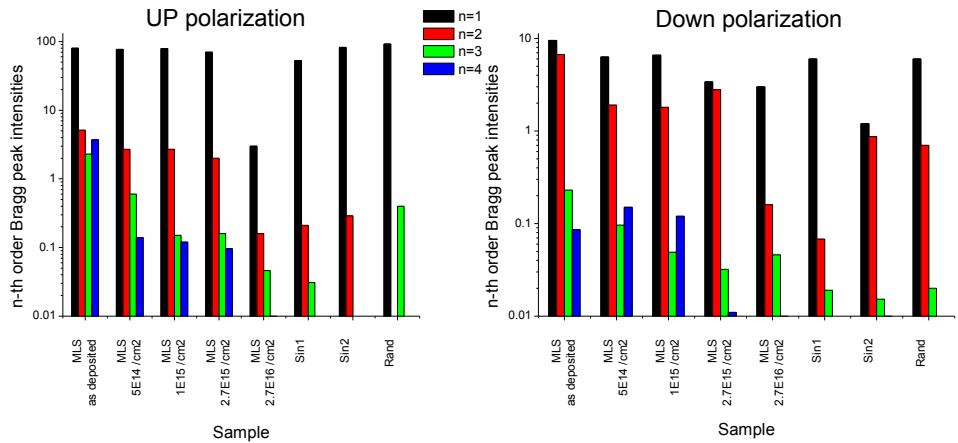

**Figure 6 Bragg** peak intensities of the various samples in this study. For details see text

Disregarding the sample irradiated with Ne$^+$ of the highest fluence, the irradiation improved the monochromatisation of the "MLS" sample, decreasing the total intensity of higher order reflections from 11.1% to 2.2% without substantially decreasing its reflection (less than of Be-filter which would reduce the intensity by about 20%). Considering all samples, the highest reflectivity (92%) is found in the case of sample "rand" followed by sample "sin2". For further consideration the polarizing efficiency and the intensity of higher harmonics are plotted in Figure 7.

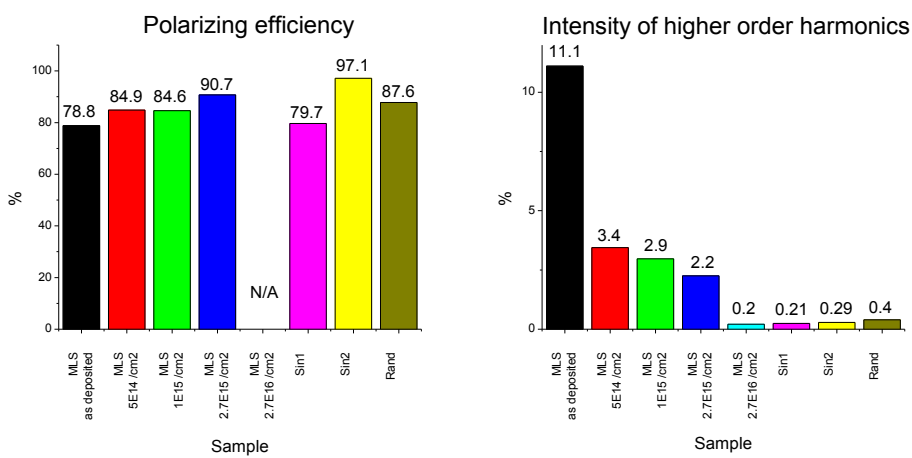

**Figure 7** Polarizing efficiency and intensity of higher order harmonics calculated from PNR measurements

According to the results it can be stated that the irradiation also improves the polarizing efficiency of the samples (from 78.8% to 90.7), however the best value corresponds to the "sin2" sample. Sample "rand" possesses a polarizing efficiency of 87.6% but it can be remarkably increased (not shown) by incorporating nitrogen in the Si layer via gas addition to the camber during Si deposition.

## 4. Summary

The neutron beams required for monochromatic applications are often contaminated by higher order intensity contributions which passed the monochromator. Three types of layer structures were prepared by DC magnetron sputtering. In the first case an approximate sinusoidal profile was to be achieved directly by carefully controlling the evaporation parameters during sputtering process. Next we introduced a random distribution of the layer thicknesses where the sample layer structure was derived from a desired simulated spectra. Finally we set out from a perfect multilayer structure and a sinusoidal profile was meant to be achieved by 0, 5, 10, 27 and $270\times10^{15}$ /cm$^2$ fluence of 350 keV neon irradiation. By neglecting the highest fluence irradiated sample (where the adjacent layers totally mixed), we observed improving of the monochromatisation of the "MLS" sample by decreasing the total intensity of higher order reflections from 11.1% to 2.2% with the intensity of the first order Bragg peak decreasing from 80% to 70%. The polarizing efficiency also increased with higher fluence of irradiation from 78.8% to 90.7%. We could achieve the highest reflectance at the "rand" sample (92%) with 87.6% polarizing efficiency. The sample with approximate layer structure produced the highest polarizing efficiency (97.1) with reflectance of 82%.

## 5. Acknowledgement

This work was partially supported by the Hungarian National Science Fund (OTKA), the National Office for Research and Technology of Hungary under contract K62272, NAP-VENEUS'05